\documentclass{article}

\usepackage{arxiv}

\usepackage[utf8]{inputenc} 
\usepackage[T1]{fontenc}    
\usepackage{hyperref}       
\usepackage{url}            
\usepackage{booktabs}       
\usepackage{amsfonts}       
\usepackage{nicefrac}       
\usepackage{microtype}      
\usepackage{multirow}
\usepackage{graphicx}
\usepackage{natbib}
\usepackage{doi}
\usepackage{float}

\title{Continuous Visual Feedback of Risk for Haptic Lateral Assistance}

\author{ \href{https://orcid.org/0000-0003-3704-1415}{\includegraphics[scale=0.06]{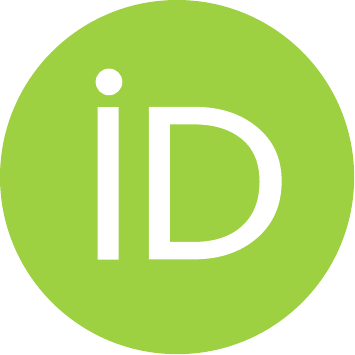}\hspace{1mm}Gyanendra Sharma} \\
	Toyota Research Institute\\
	Los Altos, CA 94022 \\
	\texttt{gyanendra.sharma@woven-planet.global} \\
	\And
	\href{https://orcid.org/0000-0003-2708-8623}{\includegraphics[scale=0.06]{pics/orcid.pdf}\hspace{1mm}Hiroshi Yasuda} \\
	Toyota Research Institute\\
	Los Altos, CA 94022 \\
	\texttt{hiroshi.yasuda@tri.global} \\
	\And
        \href{https://orcid.org/0000-0002-2852-4020}{\includegraphics[scale=0.06]{pics/orcid.pdf}\hspace{1mm}Manuel Kuehner} \\
	Toyota Research Institute\\
	Los Altos, CA 94022 \\
	\texttt{manuel.kuehner@woven-planet.global}
}

\date{}

\renewcommand{\headeright}{ }
\renewcommand{\undertitle}{ }
\renewcommand{\shorttitle}{Visual Feedback for Haptic Lateral Assistance}

\hypersetup{
pdftitle={Continuous Visual Feedback of Risk for Haptic Lateral Assistance},
pdfauthor={Gyanendra Sharma, Hiroshi Yasuda and Manuel Kuehner},
pdfkeywords={Advanced Driving Assist System, Lateral Assistance System, Shared Control, User Acceptance, Head-Up Display, Visualization},
}

\begin{document}

\maketitle

\begin{abstract}
Lateral assistance systems are a significant part of the currently existing Advanced Driving Assistance System (ADAS). In most cases, such systems provide intermittent audio and haptic feedback rather than continuous feedback, making it difficult for drivers to form a clear mental model. In this paper, we propose continuous visual feedback for the lateral assistance system by leveraging the Head-Up Display (HUD) alongside haptic feedback through the steering wheel. The HUD provides visualization of the risk profile underlying the haptic feedback. We hypothesize that our proposed visualization helps form a clear mental model and improves the system's acceptance. We conduct a user study on a simulated version of a car driving on a two-lane freeway and compare the haptic lateral assistance system with and without the visualization on the HUD. While both conditions received high acceptance scores, there was no significant gain or deterioration in acceptance between them. We discuss potential improvements in the visualization based on anticipation performance and qualitative feedback from users.
\end{abstract}

\begin{figure}[H]
\centering
      \includegraphics[width=\textwidth]{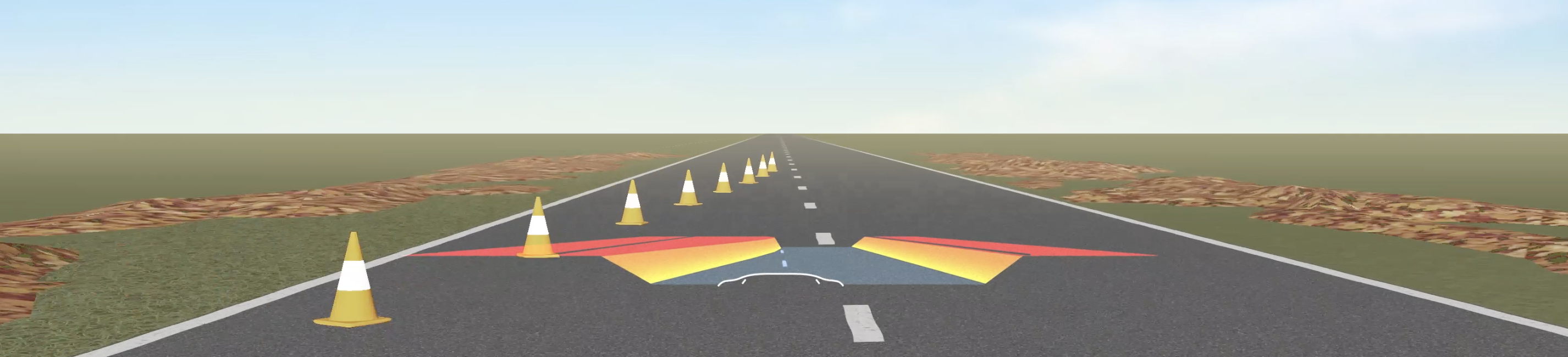}
  \caption{Visuals on the head-up display based on the underlying risk model.}
  \label{fig:teaser}
\end{figure}
\section{Introduction}
In most existing lateral assistance systems, the feedback to the user is primarily guided through the torque applied to the steering wheel along with sound. The feedback occurs intermittently and in cases where the perceived risk from immediate surroundings does not match between the driver and the vehicle, it may even appear to be arbitrary. This may result in conflict, especially in cases where both the driver and the vehicle are simultaneously providing input to the steering wheel~\cite{itoh2016hierarchical}. Continuous exposure to such situations leads to annoyance and ultimately towards drivers turning off these systems entirely~\cite{marshall2007alerts}. One of the key elements that can reduce such nuisance is clear communication that leads to a good cooperative system~\cite{petermeijer2021makes}. Recently, the addition of visual feedback to such systems, in the form of Head-Up Display (HUD) has shown promising results in providing drivers with an extra layer of information regarding vehicle perception and its comprehension of the surrounding environment~\cite{currano2021little, colley2021should}. In our work, we apply this technology to provide continuous visual feedback to the driver, with the expectation that it helps in minimizing misconceptions and annoyance and ultimately eliminate the need to turn off such systems.

Merits of visual feedback in the form of HUD have been long studied in varying driving contexts over the years~\cite{campbell2018human, gabbard2014behind, marcano2021shared}. There are many considerable advantages to utilizing the HUD as part of the Advanced Driver Assistance System (ADAS) - information provided through the HUD with well-intentioned design and placement enables drivers to receive feedback without taking their eyes off the road~\cite{park2013efficient, maroto2018head, liu2004comparison}. Kim and Dey showed that utilizing HUD to display navigation instead of an external device that takes eye gaze off the road significantly improved distraction-related measures~\cite{kim2009simulated}.

In addition to leveraging the HUD as a modality to communicate visual information, we also propose continuous visual feedback as part of the lateral assistance system in contrast to the commonly existing assistance system that relies on intermittent communication via the haptic modality i.e brake, nudges on steering wheel, etc. Such discontinuous warning systems rely on timing and this is a well-studied problem especially in the context of Forward Collision Warning (FCW)~\cite{aoki2011perceptual, kondoh2008identification}. However, in recent times, there have been investigations into how binary warning systems may not be ideal for everyone~\cite{montgomery2014age}. Significant effort has been made to explore the assistive system without binary warnings~\cite{abbink2018topology, marcano2020review}. Our work focuses on the same domain by providing users with a warning system in the form of risk visualization in HUD that is continuous and present at all times.

Lastly, we propose a visualization design that can represent risk based on the critical boundaries reported by the perceptual system of the vehicle. This risk of boundary is exactly the same that is used for haptic feedback on the steering wheel in a lateral assistance system. An illustration of how the visualization appears in the context of surrounding risk elements is shown in figure~\ref{fig:teaser}. The foundation is derived from the risk component which will be discussed in detail in the latter sections. The visual is characterized by always being "ON" but the color and shape properties of the overall visual artifact change per the vehicle's perception to continuously provide clear information to the driver. This makes the feedback from the lateral safety system truly multimodal, which has shown more promising results in communicating with the drivers than unimodal ones~\cite{politis2014evaluating}.

\section{Related Work}
The evolution of safety systems in vehicles especially due to the availability of advanced sensing and interpretation capabilities has made driving into a human-machine shared control paradigm, where the vehicle actively aids the driver in keeping its occupants safe~\cite{marcano2021shared, iihseffective}. Many 
safety systems such as the one discussed in this work; lateral assistance, rely on communicating via intermittent nudges in the steering wheel, in addition to sound instead of continuous feedback. Although effective in many cases, this has also caused annoyance leading to turning such systems off by the drivers~\cite{iihsannoyance}. Approaches, albeit in the realm of semi-autonomous systems, that focus on more continuous feedback systems have been proposed~\cite{faltaous2018design}. Studies based on other existing safety systems such as the Automatic Cruise Control (ACC) have shown that continuous feedback leads to a clearer mental model and more appropriate understanding of the system~\cite{seppelt2019keeping}. Our work follows in a similar vein to translate learning outcomes from these studies to the lateral assistance system framework by instrumenting a continuous visual feedback through HUD.

While an additional visual feedback layer may lead to an increase in cognitive load, the mode of delivery i.e the head-up display, significantly curtails such grievances~\cite{charissis2015improving, liu2004comparison}. Within the modality of the HUD, we also take special care to place the visual artifact such that drivers do not have to move their head or eye gaze away from the simulated road as suggested by Topliss et al.~\cite{topliss2019evaluating}. Similarly, HUDs allow for drivers to maintain safe driving performance for longer~\cite{hannah2020long}, and in comparison to other display modes, contribute to lower workload~\cite{horrey2003does}, and in many cases are preferred by the drivers~\cite{jose2016comparative}. Recently, in-vehicle light displays are also being studied as an alternative mode of communication in comparison to traditional mediums such as the Information Cluster (IC)~\cite{locken2016enlightening, schmidt2017guiding, van2017ambient}. However, such systems are out of scope at this time for our work.

In regards to the visualization, there has been a lot of focus on designing navigational and informational interfaces for the HUD~\cite{parsch2019designing, currano2021little}. In contrast, our work offers a novel approach in visualization design that can incorporate lateral risk information which inherently drives the haptic steering wheel feedback in a lateral assistance system. The design principles are guided with an aim to visualize risk in a continuous and contextual manner. Although our work doesn't include the usage of Augmented Reality (AR) based HUD, we remain cognizant and inspired by previous work on interface design approaches proposed more specifically for AR based in-vehicle HMI systems~\cite{merenda2018augmented}.

\section{Risk Model for Haptic and Visual Feedback}
\begin{figure}
    \centering
    \includegraphics{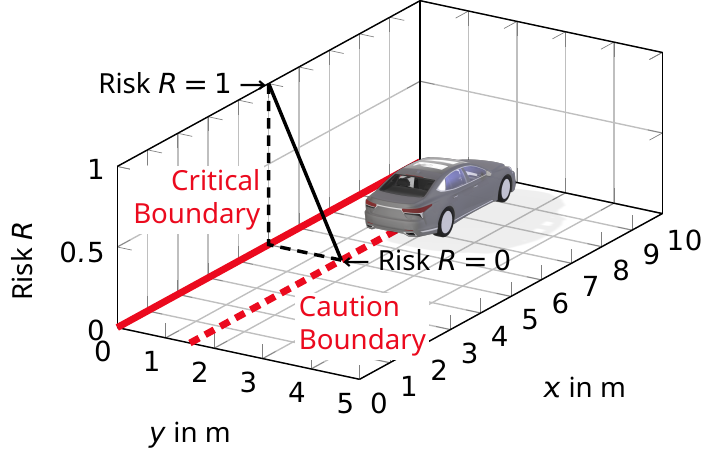}
    \caption{Depiction of the concept of a spatial risk distribution. The ADAS wants to prevent the driver from going left of the critical boundary. The risk $R$ is increasing linearly from zero (minimum) to one (maximum) within the caution boundary and the critical boundary.}
    \label{fig:MK_Fig01_PDFA}
\end{figure}
\raggedbottom
The goal of our continuous visual and haptic feedback is to aid drivers in understanding the current risk level associated with the car's position on the road relative to road elements, such as road edges, or other road users; cars, bicyclists, pedestrians, etc. Most lateral assistance system rely on lane/road markings alone and lack contextual risk information. We propose an added layer; a risk model to visualize and provide feedback on the steering wheel based on positional context.

\begin{figure}
    \centering
    \includegraphics{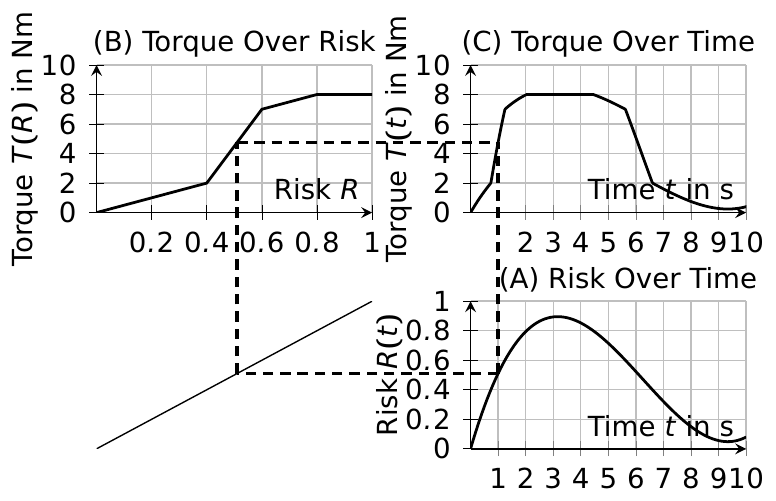}
    \caption{Logic of converting risk $R(t)$ (A) to torque $T(t)$ (C) using a risk-to-torque $T(R)$ conversion (B). See figure~\ref{fig:MK_Fig01_PDFA} for how the risk is calculated.}
    \label{fig:MK_Fig02_PDFA}
\end{figure}
\raggedbottom
For instance, consider a situation shown in figure~\ref{fig:MK_Fig01_PDFA}, where a person is driving a car on a surface in the positive $x$ direction. The lateral assistance system gets two kinds of boundaries from the overarching ADAS; a critical boundary (solid red line), and a caution boundary (dashed red line). A common example of critical boundary could be the edge of a road. The ADAS wants to prevent the driver from going left of the critical boundary. The caution boundary is spatial padding used by the lateral assistance system to communicate the proximity of the critical boundary to the driver through haptic feedback and visual feedback. As an illustration, figure~\ref{fig:MK_Fig01_PDFA} depicts the caution boundary as a constant 1.5\,m offset to the critical boundary. We assign a risk value to the position of the car when it is within the two boundaries (critical and caution). At the caution boundary, the risk is zero (\textit{R} = 0), and at the critical boundary, the risk is one (\textit{R} = 1). The risk changes linearly between these two boundaries; see the solid black line in figure~\ref{fig:MK_Fig01_PDFA}. As mentioned above, the goal of the lateral assistance system is to communicate the proximity of the critical boundary to the driver through haptic and visual feedback.

Figure~\ref{fig:MK_Fig02_PDFA} explains how our proposed system converts the abstract concept of risk into the user domain through the steering wheel's haptic feedback in the form of repelling torque feedback on the steering wheel alongside visualizing the same risk on the HUD. Figure~\ref{fig:MK_Fig02_PDFA}, diagram~A (bottom) shows the risk as a function of time $R(t)$ over 10~seconds. The risk $R(t)$ is then converted into repelling torque as a function of time $T(t)$, as seen in figure~\ref{fig:MK_Fig02_PDFA}, diagram~C (top right). The conversion between risk as well as torque as a function of time is controlled by the risk-to-torque relationship $T(R)$. Figure~\ref{fig:MK_Fig02_PDFA} shows a specific risk-to-torque relationship $T(R)$ in diagram~B (top left), however, this is only for illustrative purposes.
\begin{figure}[H]
    \centering
    \includegraphics[width=100mm]{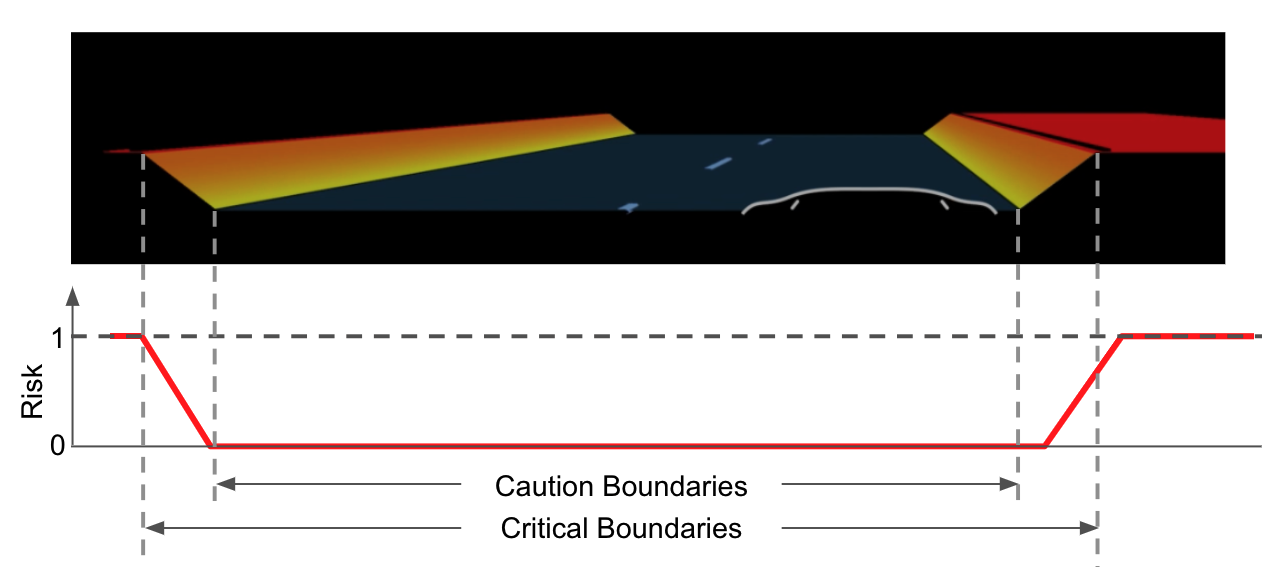}
    \caption{The visualization in the HUD based on the risk profile.}
    \label{fig:risk_visual}
\end{figure}
\raggedbottom
The visualization that appears on the HUD as shown in figure~\ref{fig:risk_visual} follows from the same concept that guides the feedback on the steering wheel. For instance, as shown in the same figure, the car (represented by a front hood) is currently in the no-risk (\textit{R} = 0) region, and well within the caution boundary. However, if it were to steer right, it would cross the caution boundary and visually enter the yellow zone, indicated by the car hood symbol entering into the yellow region. This would happen in conjunction with the feedback on the steering wheel where both the visual and haptic feedback follows the underlying risk model described earlier. For instance, if the car were to continue steering to the right even after entering the yellow zone, the increase in repellent force on the steering wheel would be accompanied by the visuals displaying the car to be in a region with an increasing gradient of red color, signifying increasing risk. In regards to the visual elements, note that the design considerations are only for a regular HUD display and do not apply to an AR-based HUD. The risk visuals are not overlaid on the road sections upfront but the animated visual includes representations of the road ahead in combination with the risk components.
\section{Methodology}
\subsection{Experimental Setup}
\begin{figure}[H]
    \centering
    \includegraphics[width=100mm]{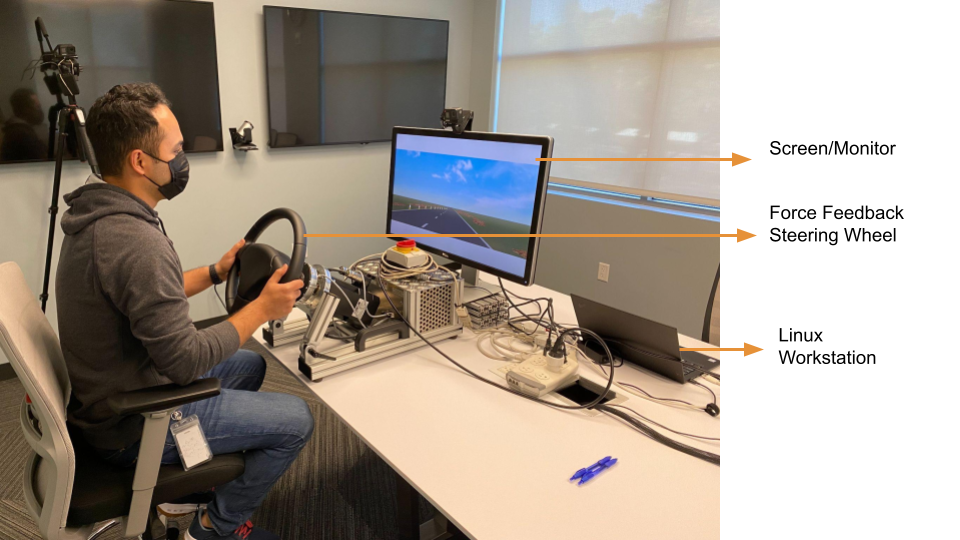}
    \caption{Setup of the driving task simulator. The participant holds the steering wheel while the virtual environment and the HUD visualization are shown on the computer screen, both of which are connected to a workstation.}
    \label{fig:setup}
\end{figure}

As shown in figure~\ref{fig:setup}, the experiment consists of a simulated tabletop driving setup. We use a 2D screen to overlay the visual aspects, which includes the HUD, road view as well as the driving scenery. The force feedback steering wheel provides the haptic feedback by applying torque in either clockwise or anti-clockwise direction, depending on the road conditions ahead. The setup is imagined to work as a Level 2 system, where the user is actively monitoring and has both hands on the wheel at all times, but the simulated car does all of the driving, including obstacle avoidance and lane change.
\raggedbottom
\subsection{Participants}
We recruited 15 (8 Female, 7 Male) participants to evaluate our proposed system. The distribution in terms of the age of the participants ranged from 23 to 55 years for all but 2 participants, who were older. All participants indicated that they had some kind of experience with the existing Advanced Driver Assisted Systems, which is quite pervasive in recent years. All participants are regular drivers with at least 5+ years of experience. Participants were required to be fully vaccinated, and show a negative test before taking part in the research as part of the research institutions' COVID-19 policy. Participants were compensated to take part in the study as well as for the COVID-19 test.
\subsection{Experimental Design}
We design a within-subject study, with and without HUD visualization, and ask the participants to evaluate their experiences. As shown in figure~\ref{fig:setup}, the experimental task required the participants to hold onto a force feedback steering wheel while observing the visuals shown on the screen. The system was designed to simulate an active steering guidance system that continuously provides haptic and visual feedback. The haptic feedback was characterized by locking the steering wheel in a position corresponding to the visuals on the screen by using a small resistance force. Instead of limiting the guidance system to lane keep assist or similar safety systems, we designed it to adapt and provide guidance based on the risk factors on the road. For instance, when the car perceives an obstacle ahead, the steering wheel would automatically rotate while the HUD showed the visuals based on our risk-based concept described in section 3. So, in cases where there were no obstacles, the steering wheel would be locked at a straight position that could be easily overcome with a force greater than 2 Nm. The goal is to provide feedback in the form of resistance force that the driver can clearly feel but isn't too strong if they wanted to take control.
\begin{figure}[H]
    \centering
    \includegraphics[width=80mm]{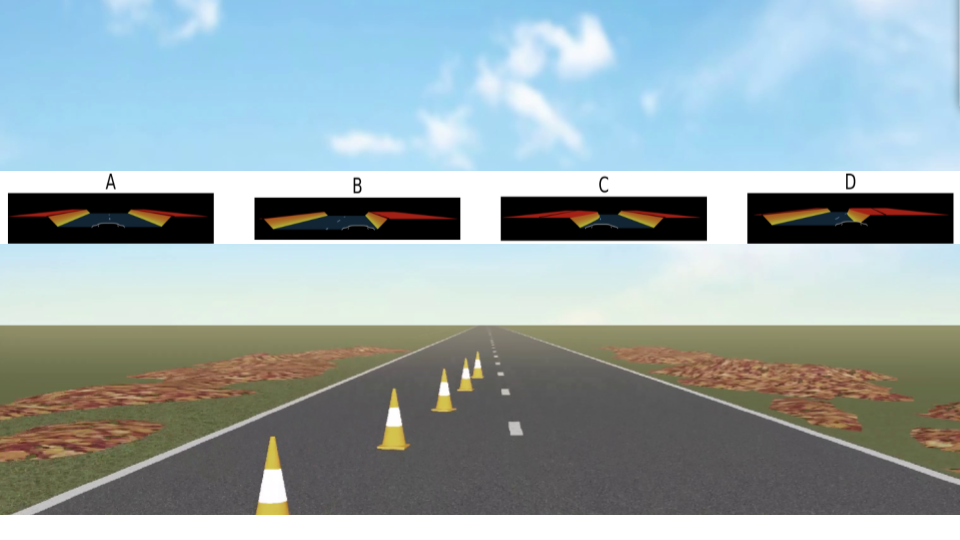}
    \caption{Answer choices for an anticipation task. For instance, in this particular case, considering the car position and the level of risk ahead, correct user response would be C.}
    \label{fig:anticipation_options}
\end{figure}
\raggedbottom

Based on our experimental design, we measure the following metrics; acceptance, physical effort, anticipation and usefulness with respect to criticality. In regards to acceptance, we use the method proposed by Van Laan et al.~\cite{van1997simple} to assess our HUD-based system acceptance on two dimensions, a usefulness scale, and an affective satisfying scale. For physical effort, we measure the Torsion Bar Torque(TBT) resulting from the effort applied by the participant on the force feedback steering wheel~\cite{abbink2018topology}. In order to obtain anticipation scores i.e to test whether participants were able to form a clear mental model of the HUD visualization as well as the haptic feedback of our proposed system, we devise an anticipation test session at the end of the experiment. As shown in figure~\ref{fig:anticipation_options}, at various sections of the video, the screen is paused. Then, participants respond in regards to which among the four options of HUD visual they anticipate seeing based on the road conditions at the paused section. They also verbally describe what they anticipate in terms of haptic feedback on the steering wheel. For each participant, there are 4 unique questions. Lastly, to obtain usefulness scores with respect to criticality, we administer a Likert scale questionnaire asking participants to rate our proposed system on the basis of their usefulness impressions in critical situations.

To avoid ordering influence, each participant was first exposed to either the HUD or no HUD condition depending on the order of exposure of the previous participant. This allowed us to collect evaluations with approximately even (8 participants with HUD condition first, 7 with the non-HUD condition first) distribution. 

\subsection{Experimental Procedure}
Once the participants are ushered into the test room after completing all required COVID protocols, they sign the IRB-approved consent form. We record the experimental sessions using video cameras after receiving verbal consent to do so from the participants. The participants are briefly explained what they are expected to do during each of the sessions.

In the first session, the participant experiences the simulated lateral assistance system with or without HUD visualization based on the order of the previous participant. The session lasts for approximately 2.5 minutes. After completion, the participant fills out the acceptance scale questionnaire for that particular condition. Then, we repeat the session with the second condition and fill out the acceptance scale questionnaire based on it. 

The third session comprises of anticipation test, where the participant answers four questions based on the freeze-frame sections as shown in figure ~\ref{fig:anticipation_options}. After this session, participants fill out a questionnaire to assess their impressions of the usefulness of the system with and without visualization in critical and non-critical situations.

\section{Results}
\subsection{Acceptance}
\begin{figure}[H]
  \centering
  \begin{minipage}[b]{0.4\textwidth}
    \includegraphics[width=60mm]{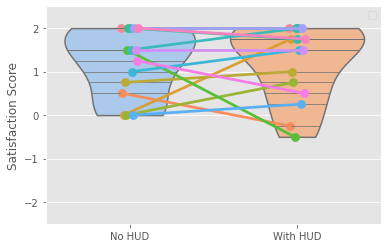}
    \caption{Satisfaction score summary. }
    \label{fig:satisfaction}
  \end{minipage}
  \hfill
  \begin{minipage}[b]{0.4\textwidth}
    \includegraphics[width=60mm]{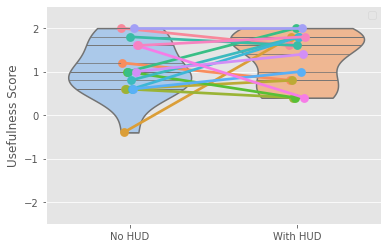}
    \caption{Usefulenss score summary.}
    \label{fig:usefulness}
  \end{minipage}
\end{figure}
\raggedbottom
\begin{table}[h]
\centering
\begin{tabular}{c|cc|cc|c}
 & \multicolumn{2}{c|}{\begin{tabular}[c]{@{}c@{}}Without\\ HUD (control)\end{tabular}} & \multicolumn{2}{c|}{\begin{tabular}[c]{@{}c@{}}With HUD \\ (treatment)\end{tabular}} & \multicolumn{1}{l}{p-value} \\ \cline{2-5}
 & \multicolumn{1}{c|}{Mean} & SD & \multicolumn{1}{c|}{Mean} & SD &  \\ \hline
Usefulness & \multicolumn{1}{c|}{1.07} & 0.65 & \multicolumn{1}{c|}{1.31} & 0.60 & 0.29 \\ \hline
Satisfaction & \multicolumn{1}{c|}{1.20} & 0.78 & \multicolumn{1}{c|}{1.20} & 0.86 & 1.00
\end{tabular}
\caption{Acceptance score summary table.}
\label{tab:acceptance}
\end{table}
\raggedbottom

In terms of acceptance scores, as shown in Table~\ref{tab:acceptance}, and Figures~\ref{fig:usefulness} and \ref{fig:satisfaction}, we see that the average score for both HUD and no HUD conditions are quite high. The mean score for the usefulness score or the treatment condition is slightly higher but not significant.
\subsection{Physical Effort}
\begin{figure}[H]
    \centering
    \includegraphics[width=80mm]{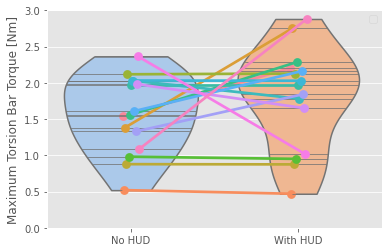}
    \caption{Maximum Torsion torque comparison between each participant.}
    \label{fig:tortiontorque}
\end{figure}

We measure the effort based on whether the participants try to overcome the resistance or the haptic feedback by measuring the Torsion Bar Torque (TBT) on the steering wheel. We look at the maximum TBT applied by each participant for each session as a metric to determine if there are any noticeable differences in effort between the two experimental conditions. The results are shown in figure~\ref{fig:tortiontorque}, where we see that there is no significant difference from two-sided paired T-test; t(14)=-1.25, \textit{p}=0.23. The experimental procedure instructs participants to hold on to the steering wheel gently. However, this metric suggests that even in cases where effort isn't required from the participants, the familiarity of holding onto a steering wheel and applying effort seems reflexive.
\raggedbottom

\subsection{Anticipation}
 \begin{table}[h]
 \centering
\begin{tabular}{c|cc|cc|c|}
 & \multicolumn{2}{c|}{Haptics} & \multicolumn{2}{c|}{\begin{tabular}[c]{@{}c@{}}HUD \\ Visuals\end{tabular}} & \multicolumn{1}{l|}{p-value} \\ \cline{2-6} 
 & \multicolumn{1}{c|}{Mean} & SD & \multicolumn{1}{c|}{Mean} & SD & \multirow{2}{*}{0.014} \\ \cline{1-5}
\multicolumn{1}{|c|}{Anticipation Score} & \multicolumn{1}{c|}{0.95} & 0.19 & \multicolumn{1}{c|}{0.72} & 0.34
\end{tabular}
\caption{Results of anticipation scores for haptics and HUD visualization.}
\label{tab:anticipation}
\end{table}

\begin{figure}[H]
    \centering
    \includegraphics[width=80mm]{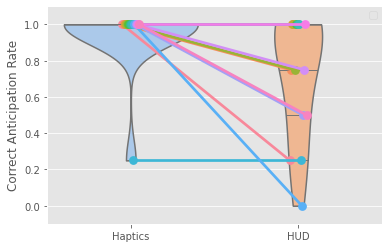}
    \caption{Anticipation test comparison between haptic and HUD visual scores for participants.}
    \label{fig:anticipation}
\end{figure}

\raggedbottom
In regards to the anticipation task, as shown in figure~\ref{fig:anticipation} and table~\ref{tab:anticipation}, participant were significantly able to better anticipate haptic feedback than the HUD visuals. However, a 72\% correct score for HUD visuals that the participants only saw for no more than 5 minutes shows promise in terms of the future steps.
\raggedbottom

\subsection{Usefulness w.r.t Criticality}
In most cases, participants agreed that continuous feedback in visual form would be very helpful in critical situations to understand what the car is trying to communicate to the driver. Participants rated the HUD based system (Mean = 1.4, SD = 0.90),  as being significantly helpful in critical situations compared to the no HUD condition (Mean = 0.4, SD = 1.29, One-sided paired t-test, \textit{p} = 0.02). This is in accordance with previous work, for instance, Colley et al. showed that in critical situations, drivers are more likely to seek information through visualization about the vehicle's intentions and interpretations of its surroundings~\cite{colley2021should}.

\subsection{Qualitative Feedback}
Participants, in describing their impression of the HUD visualization used the keywords such as "felt nice", "informative", and "helpful". One of the major points of agreement among the participants was that, having the visualization allowed them to better understand the car's comprehension of the surrounding. This is in alignment with one of our objectives, where we would like the drivers to understand the car's perception and intention in real-time through the HUD visualization. One participant also mentioned that the assistance system without visualization "didn't provide a way to anticipate or expect steering guidance inputs/actuation." This is one of the major pain points of the currently existing lateral safety systems, as actuation of nudges on the steering wheel occurs without clear communication of the underlying reasons. Du et al.~\cite{du2021designing} investigate similar issues in the context of designing alert systems in takeover transitions and show that usefulness and ease of use are dependent on communicating both the \textit{why} and \textit{what will} information. 

On the other hand, some participants suggested that the HUD visualization might be more useful if employed only during critical situations. There were concerns that the visualization being "ON" at all times was a bit distracting even though it might still be considered useful in critical situations. Few participants indicated that, they would prefer the visualization to disappear in normal driving conditions and turn "ON" only when there are obstacles or risk ahead. There were also concerns regarding the complexity of the visualization based on color combination and contrast.

\section{Discussion and Future Work}
In this work, we propose a HUD-based continuous visual feedback based on risk for lateral assistance system. In terms of acceptance, the difference between the systems with and without visualization is not significant. Based on the qualitative feedback from users, some of them considered the visualization to be fairly involved and distracting, at least in the beginning. Considering the novelty of the visualization in addition to the underlying layers of information that are being communicated continually, this reaction is not out of the ordinary. 


To form a clear mental model of visualizations, especially novel ones, requires multi-level effort~\cite{burns2020evaluate}. This suggests that new visualization paradigms take time before they can be effortlessly understood.  On the other hand, haptic feedback, primarily the movement of the steering wheel in our case, is understood intuitively, especially for people with driving experience, which is 100\% of our recruited participants. So, it is quite unsurprising that our results show significantly higher anticipation scores for steering wheel feedback in comparison to the HUD visualization. Other factors such as novelty and risk representation also contribute to this result.

Some participants' preference towards a visualization approach that turns "ON" only during critical situations, and also finding it distracting, is expected in accordance to evidence offered from the existing literature. For instance, animated visuals on the HUD might play a role in \textit{cognitive capture}, that draws entire attention to the visuals while diverting from broader contexts and surroundings~\cite{wiener1980flight, weintraub1987huds}. We posit that, there needs further investigation to find a balance where a visualization framework that is continuous but not distracting to a detrimental level can be achieved.

One of the key elements of lateral assistance systems being turned off either temporarily or permanently is due to the annoyance resulting from auditory warnings~\cite{campbell2018human, reagan2018crash, marshall2007alerts}. Ultimately, our system is part of a larger framework to design a lateral assistance system that people do not associate with annoyance. In this work, we focused only on the visual and haptic components. However, audio feedback is a key element in instigating annoyance and that remains a limitation of this study and part of our future work. 

Interactivity was another limitation of our study method where, even if the participants were able to experience the lateral assistance system, it was primarily based on a synthetically created video and didn't require driver input either in the form of acceleration or steering. The obvious next step is to allow the participants to interactively control the driving mechanism with acceleration, braking, and steering input all controlled by the drivers themselves. This would allow for a more realistic approach to evaluate our system and gain a better understanding of user perception of the system. This would also allow us to derive better quantitative metrics in terms of physical effort and other factors. Immersive video based driving simulator has shown significant promise in prototyping and creating meaningful mental models that resemble closely to driving a real car~\cite{gerber2019video}. Similar implementation would definitely be beneficial to our work as well.

Finally, our primary objective since the onset of this work was to come up with a risk-based visualization that elevates comprehension among the drivers for the lateral assistance system, leading to higher acceptance and thus less annoyance. However, based on the results of this work, we see that a direct comparison between systems with and without visualization doesn't provide enough evidence one way or another. Improving the experimental design by incorporating interactivity, sound, and also the modality in which these feedbacks are delivered i.e using a more advanced simulator or eventually testing it in a real car would also be a major improvement and within the scope of our future work.
\section{Conclusion}
The goal of this work is to propose a risk-based visualization that can be displayed in the HUD for continuous driver feedback. Our work uses this form of visualization as a means to increase driver acceptance of the lateral assistance safety system.  Our study shows that acceptance of the risk-based visualization as part of the lateral assistance system is neither better nor worse in comparison to the system without visualization. However, qualitative responses from the participants showed a high level of correct anticipation or mental model formation despite the novelty of the visual design. In addition, our work provides a framework in which the existing binary haptic lateral assistance systems can be transformed into a more continuous mechanism by extrapolating the system to use an additional layer of feedback in the visual form. The results presented in this work suggest that, especially for critical situations, using some form of risk-based visual feedback allows the drivers to understand the vehicle's interpretation of its surroundings better. Formalizing these hypotheses and testing further to validate, in addition to improving the testing modalities remains well within the scope of our future work. 
\section{Acknowledgments}
We would like to thank the entire Human Machine Interface (HMI) team at Toyota Research Institute (TRI) for their help and support throughout the project. Special thanks to Jaime Camhi, senior manager of the team, who helped with arranging the funds and resources for the work as well as provided valuable feedback throughout. We would also like to acknowledge Scott M. Harris for creating all visual assets that were used during the experiment.

\bibliographystyle{unsrtnat}
\bibliography{references}  

\end{document}